\documentclass[runningheads]{llncs}
\usepackage{graphicx}
\usepackage{xcolor}
\usepackage{listings}
\usepackage{amssymb}
\usepackage{amsmath}
\usepackage{url}
\usepackage{pbox}
\usepackage{tabularx}
\usepackage{enumitem}
\usepackage{mathabx}
\usepackage{amsfonts}
\usepackage{hyperref}
\usepackage[super]{nth}
\usepackage{fancyvrb}
\usepackage{bibnames}

\begin{document}
\title{Best Practices for Implementing FAIR Vocabularies and Ontologies on the Web}
\author{Daniel Garijo\inst{1}\orcidID{0000-0003-0454-7145} \and \\
Mar\'ia Poveda-Villal\'on\inst{2}\orcidID{0000-0003-3587-0367}}% \and
%Third Author\inst{3}\orcidID{2222--3333-4444-5555}}
%
\authorrunning{Garijo and Poveda-Villal\'on}
\titlerunning{Best Practices for FAIR Vocabularies}

\institute{Information Sciences Institute, University of Southern California \\ \email{dgarijo@isi.edu}\\ \and
Ontology Engineering Group, Universidad Polit\'ecnica de Madrid
\\ \email{mpoveda@fi.upm.es}}
\maketitle              % typeset the header of the contribution
\begin{abstract}
With the adoption of Semantic Web technologies, an increasing number of vocabularies and ontologies have been developed in different domains, ranging from Biology to Agronomy or Geosciences. However, many of these ontologies are still difficult to find, access and understand by researchers due to a lack of documentation, URI resolving issues, versioning problems, etc. In this chapter we describe guidelines and best practices for creating accessible, understandable and reusable ontologies on the Web, using standard practices and pointing to existing tools and frameworks developed by the Semantic Web community. We illustrate our guidelines with concrete examples, in order to help researchers implement these practices in their future vocabularies. 

%illustrate our experiences and best practices
%after years of making vocabularies and ontologies available in the Web, linking to standard practices when appropriate and showing examples. We hope this becomes a guideline to ontology developers to help them make their ontologies FAIR.

%\textbf{Small motivation about why the documentation and publishing is an important aspect of ontology design and implementation} 

%\textcolor{red}{MAX 16 PAGES INCL BIBLIOGRAPHY}

\end{abstract}
\section{Introduction}
In the last decade, a series of initiatives for open data, transparency and open science have led to the development of a myriad of datasets and linked Knowledge Graphs on the Web.\footnote{\url{https://lod-cloud.net/}} Ontologies and vocabularies have been developed accordingly to represent the contents of these datasets and Knowledge Graphs and help in their integration and linking. However, while significant effort has been spent on making data Findable, Accessible, Interoperable and Reusable (FAIR) \cite{wilkinson_fair_2016}, ontologies and vocabularies are often difficult to access, understand and reuse. This may be due to several reasons, including a lack of definitions of ontology classes and properties; deprecated or unavailable imported ontologies, non-resolvable ontology URIs, lack of examples and diagrams in the documentation, or having scientific publications describing an ontology without any reference to its implementation. 

The scientific community has started to acknowledge the need for ontologies to be properly documented, versioned, published and maintained following the Linked Data Principles \cite{janowicz_five_2014} and adapting the FAIR principles for data \cite{le_franc_yann_2020_3707985}. But these recommendations do not include guidelines on how to implement them for a target vocabulary. In this chapter we address this issue by describing how to make an ontology or vocabulary comply with the FAIR principles, including examples summarizing best practices from the community and our own experience; and pointing to popular existing tools and frameworks. %\textcolor{red}{6} %developed by the Semantic Web community. 

Our guidelines are aimed at ontology engineers, and therefore the paper is structured according to their ontology development processes: Section \ref{sec:URIDesign} describes several design decisions for an ontology URI (naming conventions, versioning, permanent URIs); Section \ref{sec:doc} describes how to create a documentation that is easy to reuse and understand by others (minimum metadata, diagrams to include); Section \ref{sec:publication} illustrates how to make an ontology accessible and findable in the Web; Section \ref{sec:frameworks} points to existing end-to-end frameworks that support the ontology publication process; and Section \ref{sec:conclusions} concludes our paper. We consider the design and development of an ontology out of the scope of this paper, as it has been covered by methodologies for ontology development (e.g., LOT\footnote{\url{https://lot.linkeddata.es/}} or NeOn \cite{suarez-figueroa_neon_2010}).

\section{Accessible Ontology URI Design}\label{sec:URIDesign}
Ontologies are digital artifacts, and therefore their URIs should follow the  Linked Data Principles,\footnote{\url{https://www.w3.org/DesignIssues/LinkedData.html}} and use a URI namespace under the developer's control.
The rationale is simple: only in a domain under our control we will be able to serve the right serialization of the ontology.

When creating an ontology, it is also important to think carefully about its name, prefix and URI design. Well engineered ontologies are costly to produce, and therefore they should be accessible to other potential users (e.g., by avoiding complex URIs) and  easy to differentiate from existing vocabularies.  %avoid people confusing  1) Because you don't want people to confuse it with something else. 2) Because you want to maximize its accessibility and ability to reuse. 

In this section we describe our proposed best practices for designing URIs for ontologies to make them unique, easy to remember, and easy to maintain. We acknowledge that naming conventions may be subjective, but these practices reflect our experience in ontology development over a wide range of domains (smart cities, open science, meteorology, neuroscience, etc.) and therefore provide a good starting point. We divide our guidelines over four main key points for accessible ontology URI design: how to select a name and prefix (Section \ref{sec:name}), whether to use hash or slash namespaces (Section \ref{sec:hash}), whether to use opaque URIs (Section \ref{sec:opaque}), how to incorporate semantic versioning in your ontology URI (Section \ref{sec:versioning}) and how to make your ontology URI permanent (Section \ref{sec:purl}).

For illustrating purposes, we will be using an example ontology  throughout this paper, with the following URI:  \texttt{\url{https://w3id.org/example\#}}.

%\textcolor{red}{Dani: Should the URI be w3id.org/example/ontology? It doesn't really matter that much}

%\maria{if it is an ontology I'd use https://w3id.org/def/example\# we can talk about the good practice of using def for ontologies, kos for skos and resource for data}

%\textcolor{red}{Dani: I disagree here. The /def ns is the one OnToology uses, why is it a good practice? For example, many ontologies have the /ns at the end. I wouldn't want the reader think: oh, so they are promoting their tool. Also, I add the full redirection here, and if we suggest this, then we risk people messing around with OnToology's URIs in w3id. But I am in favor of explaining that a quick way of doing this is with OnToology.}

\subsection{Ontology name and prefix}\label{sec:name}
The name and prefix of an ontology are in most cases related to its application domain. \textbf{Short} and \textbf{simple} names will help others remember your ontology easily. An extended practice is to \textbf{use acronyms} to abbreviate the title of longer ontologies, as in ``The Data Catalog Ontology''\footnote{\url{https://www.w3.org/TR/vocab-dcat-2/}} (with prefix \textbf{dcat}) or the ``Friend of a Friend Ontology''\footnote{\url{http://xmlns.com/foaf/spec/}} (with prefix \textbf{foaf}).% or the "Vocabulary of Interlinked Datasets"\footnote{\url{http://rdfs.org/ns/void}} (with prefix \textbf{void}).  

Another aspect to consider when designing the name of your ontology is to avoid overlapping with other existing vocabularies. While it is possible to overload existing prefixes by assigning them a different URI, this often confuses potential re-users that are already familiar with the state of the art. A good strategy to prevent this problem is to look for existing prefixes in common vocabulary registries such as prefix.cc,\footnote{\url{http://prefix.cc}} Linked Open Vocabularies (LOV) \cite{vandenbussche_linked_2016} or Bioportal \cite{whetzel_bioportal:_2011}. In our example ontology, by doing a quick search LOV and Bioportal, we can see that our example ontology URI has not been defined. However, the prefix ``example'' has already been registered to refer to ``example.org'', a namespace defined to create sample URIs. Therefore we decide on \textbf{exo} (derived from \emph{example ontology}) as our namespace prefix.

%\textcolor{red}{therefore we should pick a different one to be consistent}

%If you are planning on building an ontology network, 
%you may want to make them have all the same namespace, or you may want to separate them in modules. Comment the advantages and disadvantages of each.

\subsection{Hash versus slash URIs}\label{sec:hash}
When designing the URI of an ontology, it is important to determine whether the trailing element will be a hash (“\#”) or a slash (“/”). On the one hand, using a hash makes serving the ontology easier,\footnote{\url{https://www.w3.org/wiki/HashVsSlash}} as the client looking at the URI strips only the part of the URI before the hash symbol. Everything after the hash symbol is interpreted as a \emph{fragment identifier}, and may be ignored or used by browsers to refer to the right section of the HTML documentation (if available). W3C standards usually follow the hash convention, and we will follow it in our example as well.

On the other hand, using a slash allows treating each element of the ontology as a separate entity that may be described in an independent manner. This can be useful for organizational purposes when the ontology is big (thousands of classes) and returning a full serialization or rendering a single HTML documentation may be deemed too slow, showing instead each class and property separately. The Open Biomedical Ontology network\footnote{\url{http://www.obofoundry.org/}} is an example of this convention.

\subsection{Opaque URIs for classes and properties}\label{sec:opaque}
Opaque URIs obfuscate the name of classes and properties that are part of your ontology. For instance, let us assume we have an ``ExampleClassA'' in our example ontology. In order to use opaque URIs, instead of using ``ExampleClassA'' we would associate the class with an identifier such as ``EXO\_C0001'', and use it as part of its URI. This has two main advantages: First, if we decide to change the name of the class in the future, the URI of the class would not be affected by it. Second, identifiers are language agnostic. For instance, someone using another alphabet (e.g., chinese, cyrillic, etc.) would be able to refer to the same URI with the corresponding label. Examples of ontologies that follow this convention can be found in the Open Biomedical Ontologies, but it is also followed by commonly used knowledge graphs such as Wikidata \cite{vrandecic_wikidata_2014}.

The main drawback of using opaque URIs is the difficulty of interpreting properly classes and properties, which usually requires additional tooling for displaying the right labels. For this reason, we will not use them in our example.

%https://w3id.org/example#ExampleClassA

%Advantage: language. And change of labels
%Disadvantage: difficult to understand.

%There has been a long discussion regarding the usage of  vs . If you are not sure about which one is the best for your vocabulary/ontology, here is a tip: if your ontology will be huge and will be divided in many different modules, use “/”. Otherwise use “\#”. It is easier to set up and will make it easier to point to the right spot in the documentation.
 
% The advantages can be seen here: https://www.w3.org/wiki/HashVsSlash. 
 
%Returning back to the example, this is how my ontology IRI looks like:
%http://purl.org/net/wf-motifs\#
%and a sample class will be
%http://purl.org/net/wf-motifs\#Motif

%Hash is easier maybe when there aren't that many concepts, also it eases the content negotiation in KGs. Slash is better from the organizational point of view, but I honestly don't see that many advantages.

\subsection{Ontology versioning}\label{sec:versioning}
Ontologies often have multiple versions, and these should be appropriately annotated as part of the ontology metadata (e.g., with the property \emph{owl:versionIRI} %\footnote{\url{https://www.w3.org/TR/owl2-syntax/#Ontology_IRI_and_Version_IRI}} 
 and \emph{owl:versionInfo}).%\footnote{\url{https://www.w3.org/TR/owl-ref/#versionInfo-def}}). 
We recommend using \emph{semantic versioning}\footnote{\url{https://semver.org/}} as a guideline for identifying the different versions of an ontology, as it has become a common practice in software engineering. In semantic versioning, each version id should follow the format \textbf{X.Y.Z}, where \textbf{X} represents a major version (e.g., defining a set of classes and properties to support new use cases), \textbf{Y} represents a minor version (e.g., adding a single property or class), and \textbf{Z} represents patches or quick bug fixes (updating a label, adding examples, etc.). In our example ontology, the first version is \textbf{1.0.0} (first major release), with the IRI ``\textbf{https://w3id.org/example/1.0.0}'', which we would represent in the ontology as follows:

\begin{Verbatim}[frame=single,fontsize=\scriptsize]
<https://w3id.org/example> rdf:type owl:Ontology ;
    owl:versionIRI <https://w3id.org/example/1.0.0> ;
    owl:versionInfo "1.0.0"@en .
\end{Verbatim}

As shown in the example, the version IRI is independent from the URI of the ontology (``https://w3id.org/example\#''). It is \textbf{discouraged} to  include version numbers as part of the ontology URI, as it would deeply affect interoperability of its instances. For example, consider we had used ``https://w3id.org/example/1.0.0\#'' as the ontology URI, and we had populated a Knowledge Graph with two instances of ``ExampleClassA'':

\begin{Verbatim}[frame=single,fontsize=\scriptsize]
@prefix exo: <https://w3id.org/example/1.0.0#> .
@prefix ex-inst: <https://example.org/instance/> .
ex-inst:instance1 a exo:ExampleClassA .
ex-inst:instance2 a exo:ExampleClassA .
\end{Verbatim}

If we now released another version of the ontology (1.0.1), all the URIs of ExampleClassA would change (highlighted in blue below): 

\begin{Verbatim}[frame=single,commandchars=\\\{\},fontsize=\scriptsize]
@prefix \textcolor{blue}{exo2: <https://w3id.org/example/1.0.1#> .}
@prefix ex-inst: <https://example.org/instance/> .
ex-inst:instance1 a \textcolor{blue}{exo2}:ExampleClassA .
ex-inst:instance2 a \textcolor{blue}{exo2}:ExampleClassA .
\end{Verbatim}

This is an undesired behavior, because it makes \textbf{ex-inst:instance1} and \textbf{ex-inst:instance2} instances of two different classes (\textbf{exo:ExampleClassA} and \textbf{exo2:exampleClassA}) instead of a single one. Instead, we want all the instances of a class to be compatible across different ontology versions: 

\begin{Verbatim}[frame=single,fontsize=\scriptsize]
@prefix exo: <https://w3id.org/example#> .
@prefix ex-inst: <https://example.org/instance/> .
ex-inst:instance1 a exo:ExampleClassA .
ex-inst:instance2 a exo:ExampleClassA .
\end{Verbatim}

By following this convention, we can continue doing ontology releases with appropriate versioning, while keeping the classes and properties consistent in terms of their URI.

%and these should be appropriately captured in their \emph{version IRI}. When importing an ontology into another it is key to identify the version of the ontology being imported, in case that version changes in the future. 

%The version of the ontology should not be captured in the main ontology URI, as because otherwise each version would be a new ontology and it would be chaotic. 

%Here say how the local versioning should be used. And how the different versions should be used when referring ontologies. This is critical in vocabularies that are under development

%First version vs latest version, etc. 
%Example adding one version into another. Software engineering best practices () indicate that you should use three numbers to refer to ontologies: Major.Minor.Patch. Major is for X. Minor is for Y. Patch for Z. In our example ontology, we released the first version under 1.0.0:

\subsection{Using permanent URIs}\label{sec:purl}
When publishing an ontology on the Web, it is recommended to think about its long term sustainability, specifically if it gets widely adopted. For example, what will happen to the domain used for the namespace URI of an ontology after several years? (i.e., when the funding for the related research project is over). Likewise, if the ontology is hosted on a server in a university or company, what will happen if the server name changes; or if the person in charge of maintaing it needs to move the ontology hosting to another institution?

Permanent URIs services are community driven initiatives designed to address these issues. The idea behind them is simple: instead of minting a URI for a resource, users may use these services to create a \emph{proxy} URI which can then be redirected to wherever the target resource is stored at any point in time. That way, if the target resource is moved, developers just have to update its location without changing its permanent URL.  There are several open, free services to create permanent URLs on the Web, among which  purl.org\footnote{\url{https://archive.org/services/purl/}} (now hosted by the Internet Archive) and w3id\footnote{\url{https://w3id.org/}} (created by the W3C Permanent Identifier Community Group and supported by several companies) are perhaps the most commonly used.
We recommend using permanent URIs in ontologies in order to support their long term sustainability. In fact, our example ontology uses a w3id:  \textbf{https://w3id.org/example}. Creating a w3id is as simple as forking a GitHub repository and following the instructions in the readme file.\footnote{\url{https://github.com/perma-id/w3id.org}} An advantage of w3id versus \emph{purl.org} is that you have control on how to redirect the ontology to its different serializations (a full example is available in Section \ref{sec:publication}).

%The URI you choose for your ontology should be permanent and defined in a domain you control. The rationale behind this is simple: imagine that somebody is reusing the concepts defined in your ontology and you change its URI. The person reusing your ontology will no longer know the proper definitions and semantics of the reused term.

%Since I assume that most of the people reading this are not willing to pay for a new domain each time a new ontology is published, I recommend defining the URI of your vocabularies/ontologies in http://purl.org. PURL stands for “persistent uniform resource locator”, and they are widely used to give persistent URIs to resources. Once you register in the page, the process is really simple. You define a new domain, wait for the approval and create the URI for your ontology. In my case it is: http://purl.org/net/wf-motifs. 

%Note 1: If you create the name under the /net/ domain things will go faster, since it is the default domain. Otherwise they’ll have to approve the domain AND the name of your vocabulary/ontology.
%Note 2: Someone could argue that by speaking to the system admin of your enterprise/university you can obtain the vocabulary URI as well. However, depending on who you are and the ontology you are working on, the URI they suggest could be something like: http://mayor2.dia.fi.upm.es/oeg-upm/files/dgarijo/wf-motifs. This is perfectly fine, but this looks more like the place where my .owl will finally be stored. If my file has to be moved, my URI will change. Using purl ensures the URI will be permanent, and that I have control over it.

\section{Generating Reusable Ontology Documentation}\label{sec:doc}

%\maria{podriamos tener una breve intro a las 3 subsecciones siguientes: metadata, HTML generation y visualization. Así quedan las 3 al mismo nivel, me parece mas homogeneo pero es un cambio menor}\textcolor{red}{Ok}

We refer to ``ontology documentation'' as the collection of documents and explanatory comments generated during the entire ontology building process \cite{suarez-figueroa_neon_2010}. These documents are critical for helping others understand an ontology: documentation provides context, accurate definitions and examples of the different concepts that are included. In fact, an important part of the documentation is usually provided within the ontology itself through ontology metadata and natural language annotations. However, some of the documents may be external to the ontology, such as the ontology requirements document, other sources used during the knowledge acquisition phase, the conceptualization diagrams, examples of use, etc. 

In this section we describe our recommended best practices to generate ontology metadata and human oriented documentation, including diagramming guidelines to show the relationships between classes in a visual manner. 

\subsection{Ontology Metadata} \label{sec:meta}
When creating an ontology, it is crucial to describe it with appropriate metadata for others to understand it correctly. For example, if some of the classes are ambiguously defined, other researchers may misinterpret their meaning when incorporating them into their work. We distinguish two main categories of metadata in an ontology: the metadata associated with the ontology itself and the metadata associated with its elements (classes, object properties, datatype properties and individuals). 

The metadata associated to the ontology itself is important to provide an overview and identify an ontology, understand its usage conditions and understand its provenance. 
Table \ref{tab:recommendedProp} shows our recommended and optional annotation properties for describing ontologies, along with candidate properties that can be reused from existing vocabularies and standards.\footnote{Other vocabularies (e.g.,  \url{https://schema.org}) are alternatives to the ones proposed. See \url{https://w3id.org/widoco/bestPractices} for additional suggestions.} The \emph{recommended} properties are license (critical for others to know how the ontology may be used; we recommend a CC-BY license); \emph{creator}, \emph{contributor}, \emph{creation date} and \emph{previous version} to track the provenance of the ontology and compare against earlier versions; \emph{namespace URI} and  \emph{version IRI} to properly identify the ontology; and \emph{prefix},  \emph{title} and  \emph{description} to provide a quick overview on what the ontology does and how to properly refer to it. Finally, a \emph{citation} is recommended for letting other users know how to attribute the authors of the ontology. For the rest of the section, we will be using the following namespaces:\footnote{For reference, the TTL version of our example ontology is available at \url{https://dgarijo.github.io/example/release/1.0.1/ontology.ttl}}

\newcommand{\URL}[1]{\url{\detokenize{#1}}}
\begin{Verbatim}[frame=single,fontsize=\scriptsize,commandchars=\\\{\}]
rdfs    <\URL{http://www.w3.org/2000/01/rdf-schema#}>
owl     <\URL{http://www.w3.org/2002/07/owl#}>
bibo    <\URL{http://purl.org/ontology/bibo/}>
foaf    <\URL{http://xmlns.com/foaf/0.1/}>
dcterms <\URL{http://purl.org/dc/terms/}>
vaem    <\URL{http://www.linkedmodel.org/schema/vaem}>
vann    <\URL{http://purl.org/vocab/vann/}>
sw      <\URL{http://www.w3.org/2003/06/sw-vocab-status/ns#}>
\end{Verbatim}

The \emph{optional} properties included in Table \ref{tab:recommendedProp} are not critical to identify or reuse a target ontology, but provide additional insight and ease its understanding. These properties include having an \emph{abstract} and \emph{see also} with an additional overview of the ontology and links to related resources; a \emph{status} to describe its maturity (first draft, specification, etc.); information about the \emph{backward compatibility} or other \emph{incompatible} versions of the ontology; the \emph{modification} and \emph{issue} dates; the \emph{original source} that led to the definition of the ontology (requirement document, use cases, etc.); information about the \emph{publisher} organization supporting the ontology; the \emph{DOI} to accessed to a publication about the ontology; and information about the \emph{logo} and \emph{diagrams} that can be used as a visual aid for the ontology.

\begin{table}[t]
\caption{Recommended and optional metadata for describing ontologies}
\label{tab:recommendedProp}
\resizebox{\textwidth}{!}{%
\begin{tabular}{|l|l|l|l|} \hline
\textbf{Property name} & \textbf{Annotation Property} & \textbf{Rationale}  & \textbf{Guideline}\\ \hline
  License     &     dcterms:license        & Usage conditions & Recommended     \\ \hline
  Creator     &     dcterms:creator       & Provenance and attribution & Recommended \\ \hline
  Contributor  &    dcterms:contributor   & Provenance and attribution & Recommended \\ \hline
  Creation date     & dcterms:created    & Provenance  & Recommended                 \\ \hline
  Previous version     & owl:priorVersion  & Provenance and comparison & Recommended \\ \hline
  Namespace URI    &  vann:preferredNamespaceUri & Identifying the ontology & Recommended \\ \hline
  Version IRI     &    owl:versionIRI        & Versioning & Recommended \\ \hline
  Prefix     &    vann:preferredNamespacePrefix & Identifying the ontology & Recommended \\ \hline
  Title   &    dcterms:title        & Understanding & Recommended \\ \hline
  Description     &    dcterms:description        &    Understanding & Recommended \\ \hline
  Citation      & 	dcterms:bibliographicCitation &  Credit & Recommended \\ \hline
  
  Abstract &  dcterms:abstract & Additional information & Optional \\ \hline
   See also & rdfs:seeAlso & Additional information & Optional  \\ \hline
   Status &  sw:status & Maturity information & Optional   \\ \hline
   Backward compatibility & owl:backwardCompatibility & Version compatibility         & Optional               \\ \hline
   Incompatibility      & 	owl:incompatibleWith &  Version compatibility  & Optional   \\ \hline
   Modification Date      &  dcterms:modified  &  Provenance and timeliness & Optional  \\ \hline
   Issued date      &  dcterms:issued          &   Provenance and timeliness   & Optional   \\ \hline
   Source      & dcterms:source    &    Provenance & Optional  \\ \hline
   Publisher      &   dcterms:published         & Provenance   & Optional                      \\ \hline
   DOI      &     bibo:doi       &    Bibliographic information  & Optional  \\ \hline
   Logo      &    foaf:logo    &  Identifying the ontology   & Optional  \\ \hline
   Diagram      &  foaf:depiction          &    Visual documentation  & Optional   \\ \hline
\end{tabular}
}

\end{table}

Table \ref{tab:recommendedClass} shows the recommended and optional metadata properties for classes, properties, data properties and individuals. Recommended metadata properties include a human-readable \emph{label} to identify the term (using as many languages as needed); and a \emph{definition} for the ontology term that is as accurate as possible. Definitions should be clear and illustrative, as classes and property names may have different meanings to different researchers.

The rest of the properties in Table \ref{tab:recommendedClass} are nice to have to improve the understanding of ontology terms. These include \emph{examples} that illustrate how to use a term; its \emph{status} (e.g., deprecated, under discussion, etc.); the \emph{rationale} for including a term in the ontology (which may reflect consensus from a discussion); and the \emph{source} material that motivated the inclusion of the term.

\subsection{Creating a Human-Readable Documentation}

\begin{table}[t]
\centering
\caption{Recommended and optional properties for describing ontology terms}
\label{tab:recommendedClass}
%\resizebox{\textwidth}{!}{%
\begin{tabular}{|l|l|l|l|} \hline
\textbf{Property name} & \textbf{Annotation Property} & \textbf{Rationale} & \textbf{Guideline} \\ \hline
   Label    &  rdfs:label           & Readibility   & Recommended                      \\ \hline
    Definition     & rdfs:comment & Understanding  & Recommended\\ \hline
Example    &  vann:example           & Understanding  & Optional                      \\ \hline
    Status     &     sw:term\_status        &   Understanding   & Optional                  \\ \hline
    Rationale     &     vaem:rationale       &  Understanding  & Optional                    \\ \hline
    Source     &    dcterms:source        &     Provenance     & Optional              \\ \hline
\end{tabular}
%}
\end{table}

Ontologies are usually designed in editors and then exported in formats (Turtle, RDF/XML, JSON-LD, etc.) that are difficult to navigate by humans.
Researchers often address this issue by pointing to an associated paper or report, but these usually describe a scientific contribution rather than the definitions of each ontology concept in detail. A better solution is to create a documentation for all the terms in the ontology. Since this can be a time-consuming task, the Semantic Web community has developed tools to help ontology documentation. Given an OWL file as input, these tools generate an HTML documentation from the metadata included in the ontology itself (by retrieving the annotation properties recommended in Section \ref{sec:meta}), creating sections for all classes, properties, data properties and named individuals. Popular tools for ontology documentation include WIDOCO \cite{damato_widoco:_2017} (an evolution of the Live OWL Documentation Environment \cite{lode} that includes automated visualization diagrams through the WebVowl tool \cite{WebVOWL}), Parrot \cite{alonso_current_2012} or OwlDoc\footnote{\url{https://protegewiki.stanford.edu/wiki/OWLDoc}} (integrated with the Prot\'{e}g\'{e} Ontology editor \cite{musen_protege_2015}).

We recommend creating an HTML documentation for ontologies, as it makes them easier for others to understand and navigate on the Web. The tools introduced above are a good starting point for documentation generation, but we recommend expanding their results with additional motivation and context for the target ontology, pointers to the requirements and rationale; and custom diagrams with examples that illustrate how to use ontologies in practice.

\subsection{Ontology visualization}
Graphical representations of ontologies help users understand their structure, relationships and usage. Since there is no standard convention for ontology diagrams, developers have adopted several approaches, such as UML-alike diagrams in the SAREF ontologies;\footnote{See the SAREF4AGRI extension \url{https://w3id.org/def/saref4agri}} semantic network oriented diagrams in the SSN Ontology;\footnote{Semantic Sensor Network Ontology \url{https://www.w3.org/TR/vocab-ssn/}} or custom diagrams as in the Provenance Ontology.\footnote{PROV-O: The PROV Ontology \url{http://www.w3.org/TR/prov-o/}}

%\textcolor{red}{Maria, we should add a footnote like: full spec available at...}
 %\cite{Gangemi_ditto:diagrams} , 
 In the last years, conventions for ontology diagrams have been proposed (e.g., VOWL %\footnote{Specification available at \url{http://purl.org/vowl/spec}}  
 \cite{VOWL2} and Graffoo\footnote{Specification available at \url{https://essepuntato.it/graffoo/specification}})
 but none have been standardized yet. 
 In this section we suggest guidelines for generating ontology diagrams based on the UML\_Ont profile proposed at \cite{haase2009d}.\footnote{The original UML\_Ont profile uses custom labels and dependencies to cover OWL 1 constructs. Here labels are mapped to the OWL and RDF(S) constructs.} The rationale for our recommendation is that UML is commonly used in software engineering, and it is familiar to software developers. 
 %In the rest of this section the proposed graphical representations will be described in the enumerations below and shown in corresponding figures, namely Figure~\ref{fig:notationClass}, Figure~\ref{fig:notationOP}, Figure~\ref{fig:notationDP}, and Figure~\ref{fig:notationInd}. For the sake of readability, specific element notations in the mentioned figures are labelled in correspondence to the enumeration items listed below:

\begin{figure}[t!]
\centering
\includegraphics[scale=0.49]{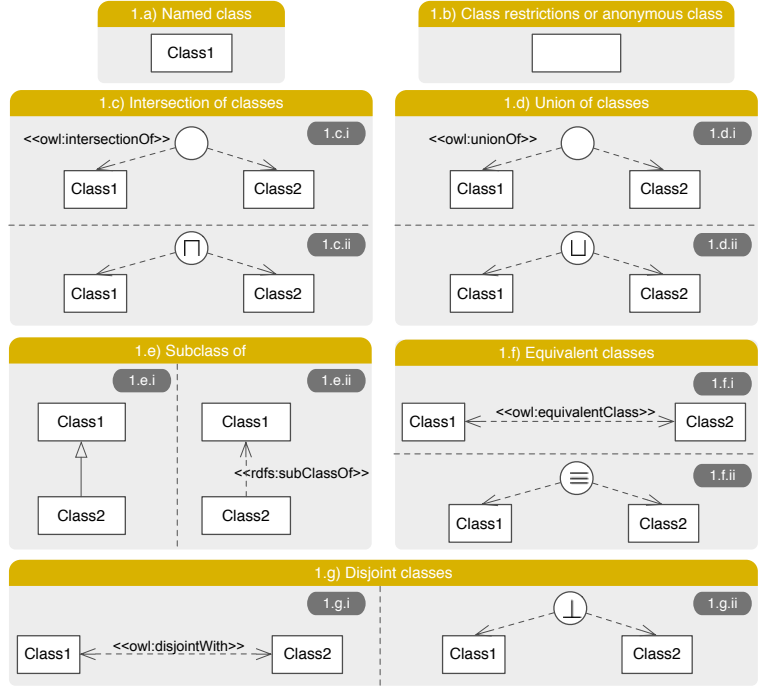}
\caption{Recommended notation for classes.}
\label{fig:notationClass}
\end{figure}

Figure~\ref{fig:notationClass} depicts our proposed graphical representations for classes, class restrictions and class axioms. \textbf{Named classes} are represented by labelled boxes (1.a); while \textbf{class restrictions or anonymous classes} are represented by empty boxes (1.b). \textbf{Intersection class} descriptions are represented either by using an empty circle with the \texttt{<<owl:intersectionOf>>} stereotype (1.c.i); or an icon including the symbol ``$\sqcap$'' (1.c.ii). Similarly, union class descriptions may use an empty circle with the \texttt{ <<owl:unionOf>>}  stereotype (1.d.i); or an icon including the symbol ``$\sqcup$''(1.d.ii). 

\textbf{Subclasses} are represented using the generalization arrow used in UML (1.e.i) or with a dependency arrow with the \texttt{<<rdfs:subClassOf>>} stereotype (1.e.ii); and \textbf{equivalent classes} are represented with double-sided UML dependency arrows with the \texttt{<<owl:equivalentClass>>} (1.f.i) stereotype or by a circle including the symbol ``$\equiv$'' (1.f.ii). Lastly, \textbf{disjoint classes} may be represented with double-sided UML dependency arrows, using the \texttt{<<owl:disjointWith>>} stereotype (1.g.i) or with a circle including the symbol ``$\bot$'' (1.g.ii).

\begin{figure}[ht!]
\centering
\includegraphics[scale=0.49]{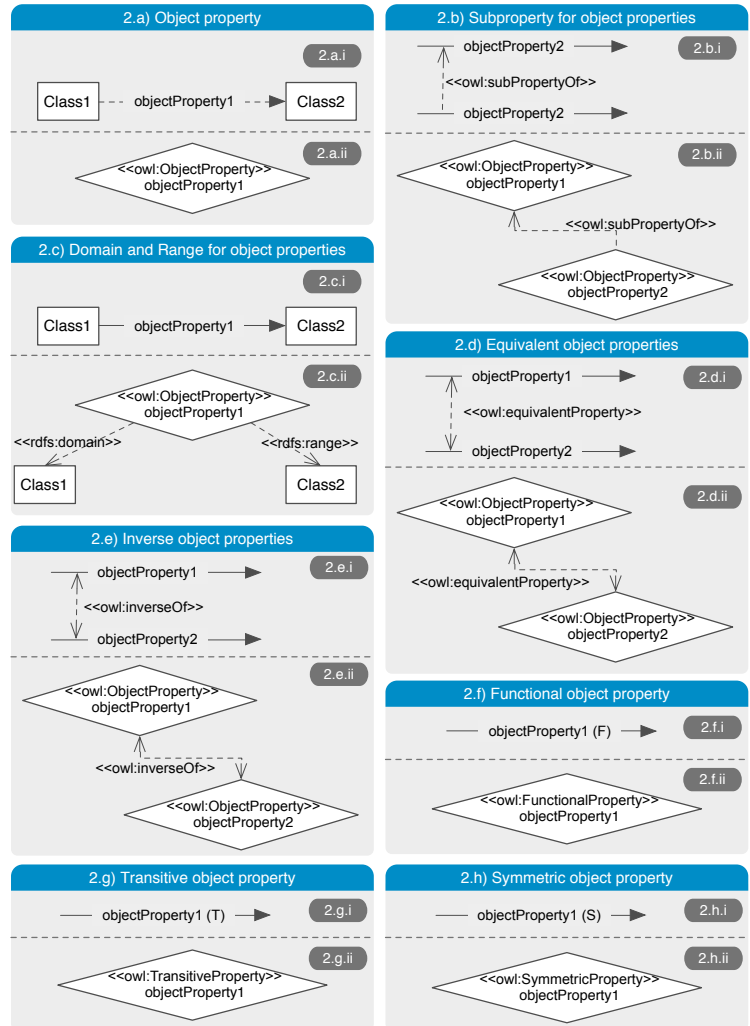}
\caption{Recommended notation for object properties.}
\label{fig:notationOP}
\end{figure}

Figure~\ref{fig:notationOP} illustrates guidelines on how to represent \textbf{object properties}. When the domain or range are not known (2.a), properties can be represented with dotted arrows (2.a.i); or with a diamond with the \texttt{<<owl:ObjectProperty>>} stereotype (2.a.ii). \textbf{Subproperties} may be represented with the UML dependency arrow with the \texttt{<<owl:subPropertyOf>>} stereotype linking the arrows that represent the involved object properties (2.b.i) or with an UML  dependency unidirectional arrow with the \texttt{<<owl:subPropertyOf>>} stereotype linking the diamonds that represent the involved object properties (2.b.ii). When \textbf{domain} and \textbf{range} are known, properties can be represented with a solid line between source and target classes (2.c.i) or with a labelled diamond accompanied by dotted arrows labelled with \texttt{<<rdfs:domain>>} and \texttt{<<rdfs:range>>} respectively (2.c.ii). \textbf{Equivalent} (2.d) and \textbf{inverse} object properties (2.e) can be represented by using a bidirectional arrow with the \texttt{<<owl:equivalentProperty>>} and \texttt{<<owl:inverseOf>>} stereotypes between the lines (2.d.i and 2.e.i) or using diamond shapes (2.d.ii and 2.e.ii). Lastly, \textbf{functional} (2.f), \textbf{transitive} (2.g) and \textbf{symmetric} (2.h) object properties can be represented using a shared notation: either by adding the first initial of the property type (F, T or S) to the object property label attached to the arrow that represents the object property; or by a labelled diamond, which represents the object property itself, including the corresponding stereotype (e.g., \texttt{<<owl:FunctionalProperty>>}).

\begin{figure}[t!]
\centering
\includegraphics[scale=0.49]{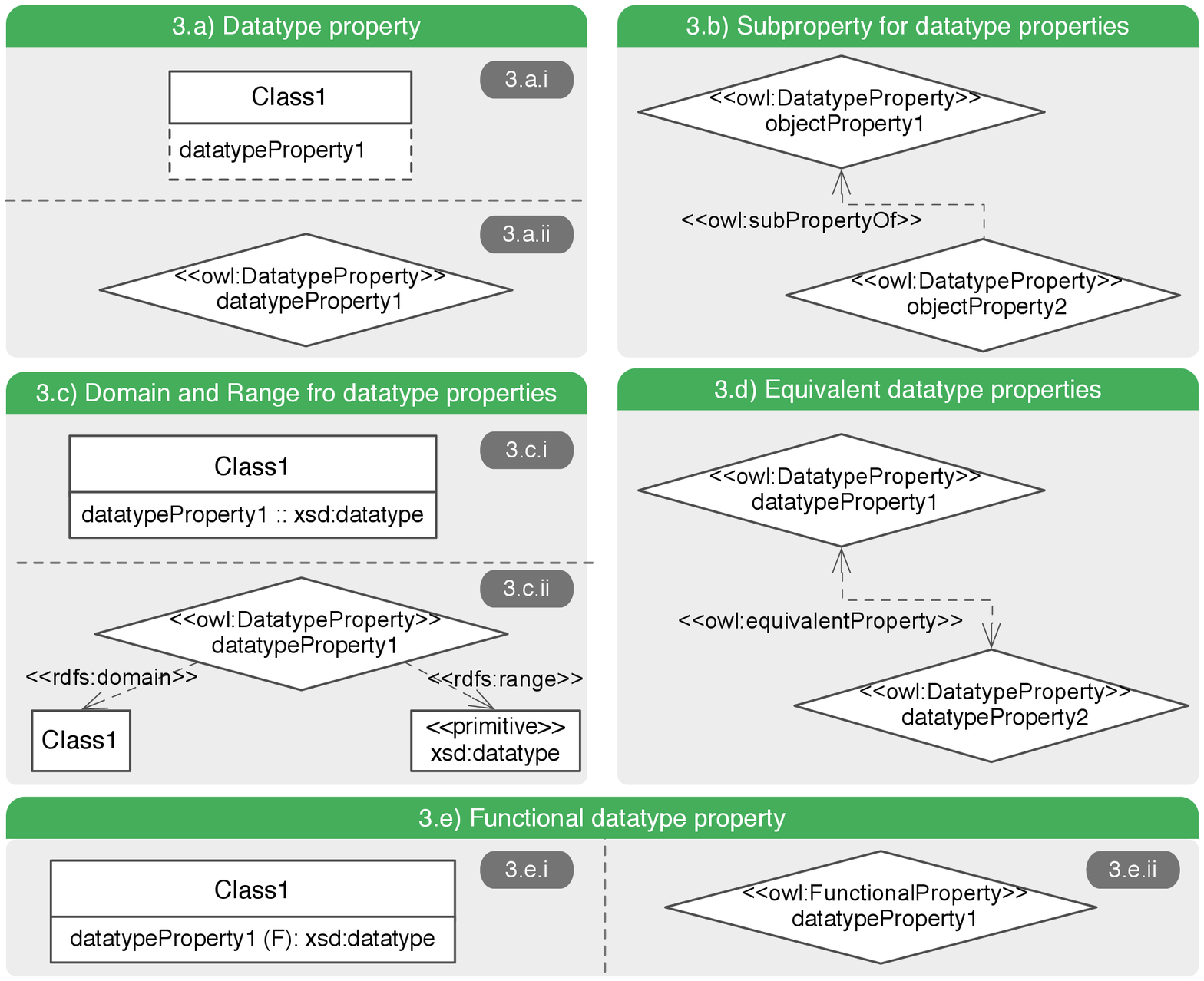}
\caption{Recommended notation for datatype properties. }
\label{fig:notationDP}
\end{figure}

Figure~\ref{fig:notationDP} shows how to represent \textbf{datatype properties}. When the domain or range are not known (3.a) datatype properties may be represented as labelled dashed boxes attached to boxes representing classes (3.a.i); or as a diamond with  \texttt{<<owl:DatatypeProperty>>}. \textbf{Subproperties} for datatypes may be represented with a UML dependency arrow with the \texttt{<<owl:subPropertyOf>>} stereotype linking the diamonds that represent the involved datatype properties (3.b). When the \textbf{domain} and/or \textbf{range} are known, the box representing the datatype property may be depicted with a solid line indicating that the domain of the datatype property is the attached class and the range may be included following the characters ``::'' after the datatype label (3.c.i). Alternatively, a labelled diamond may be used accompanied by dotted arrows labelled with the \texttt{<<rdfs:domain>>} and \texttt{<<rdfs:range>>} stereotypes respectively (3.c.ii). \textbf{Equivalent datatype properties} may be represented by a UML dependency bidirectional arrow with the \texttt{<<owl:EquivalentProperty>>} stereotype linking the diamonds that represent the datatype properties (3.d). \textbf{Functional datatype properties} may be represented by adding ``(F)'' to the datatype property label (3.e.i) or by a labelled diamond including the \texttt{<<owl:FunctionalProperty>>} stereotype (3.e.ii).

\begin{figure}[t!]
\centering
\includegraphics[scale=0.49]{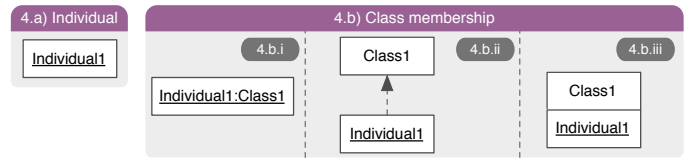}
\caption{Recommended notation for individuals. }
\label{fig:notationInd}
\end{figure}

Finally, Figure~\ref{fig:notationInd} proposes how to represent individuals and class membership. \textbf{Individuals} or instances may be represented by labelled boxes with underlined names or identifiers (4.a). {Class membership} may be represented with labelled boxes containing the individual name followed by the character ``:'' and the class name, all underlined (4.b.i); by a linking the individual box with the class using a unidirectional UML dependency arrow with the stereotype \texttt{<<rdf:type>>} (4.b.ii) or with a underlined labelled box for the individual attached to the class.

%\textcolor{red}{Creo que esta seccion es quizas demasiado larga, y diluye el mensaje del paper, que son practicas mas generales}

%\maria{Y si llamamos a esta seccion Ontology Publication y lueego la dividimos en 1) publicar distintos formatos y 2) promocion, es decir ponerla disponible en registros generales como LOV y especificos del tema de la ontologia? eso seria subir el 5.1 aqui}
%\textcolor{red}{Entonces tenemos que cambiar el titulo del paper, ya que hemos puesto publicar, pero en realidad estamos haciendo tambien disenyo.}

\section{Ontology Publication on the Web}\label{sec:publication}
Once an ontology is fully implemented and documented, it is time to make it accessible and findable in the Web. In this section we briefly describe the best practices to perform content negotiation to serve a target ontology in multiple formats (Section \ref{sec:cn}) and registries for describing new  ontologies (Section \ref{sec:findable}).

\subsection{Ontology Accessibility in Multiple Interoperable Formats}\label{sec:cn}
Ontologies should be made available in both human and machine readable manner using a single identifier: the ontology URI. This way we can make any ontology resolve to its HTML documentation when accessed by a user in a browser; and resolve to a standard RDF serialization when importing it in an ontology editor. In order to distinguish the target resource to serve (HTML or RDF serialization), we must implement a \emph{303 See Other redirect},\footnote{\url{https://tools.ietf.org/html/rfc7231\#page-57}} a common practice in the Semantic Web community for doing \emph{content negotiaiton} over URIs. This type of redirect indicates the location of a target resource in the server based on the received request, but has to be appropriately configured on the server where we are hosting the ontology. Fortunately, there are W3C best practices (recipes) on how to configure an Apache HTTP server -a commonly used type of server for serving files-  for hash ended and slash ended ontologies.\footnote{\url{http://www.w3.org/TR/swbp-vocab-pub/}} Here we expand these practices with our example ontology to illustrate: 1) How to support multiple serializations of an ontology (HTML and Turtle); 2) how to support version redirection, as we would like all the versions of the ontology to be appropriately available, not only the latest; 3) How to specify if a serialization is not supported (for example, requests for JSON-LD will return a 406 non-acceptable code, rather than an RDF/XML serialization); and 4) how to implement a default response in case the user agent doing the request does not specify a target in the request (by default we return Turtle). The \emph{htaccess} file to be placed in the server (in an \textbf{``/example''} folder)  would look as follows:

\begin{Verbatim}[frame=single, fontsize=\scriptsize]
# Turn off MultiViews (Apache-specific command)
Options -MultiViews

# Directive to specify supported types besides html and xml
# Standard RDF serialization formats include Turtle, RDF/XML, N-Triples and JSON-LD
AddType text/turtle .ttl
RewriteEngine on

# Rewrite rule for accessing the latest version.
RewriteCond %{HTTP_ACCEPT} !application/rdf\+xml.*(text/html|application/xhtml\+xml)
RewriteCond %{HTTP_ACCEPT} text/html [OR]
RewriteCond %{HTTP_ACCEPT} application/xhtml\+xml [OR]
RewriteCond %{HTTP_USER_AGENT} ^Mozilla/.*
RewriteRule ^$ https://dgarijo.github.io/example/release/1.0.1/index-en.html [R=303,L]

# Rewrite rule to serve the Turtle serialization from the vocabulary URI (latest version)
RewriteCond %{HTTP_ACCEPT} text/turtle [OR]
RewriteCond %{HTTP_ACCEPT} text/\* [OR]
RewriteCond %{HTTP_ACCEPT} \*/turtle 
RewriteRule ^$ https://dgarijo.github.io/example/release/1.0.1/ontology.ttl [R=303,L]

# Rewrite rules for retrieving a particular version.
RewriteCond %{HTTP_ACCEPT} !application/rdf\+xml.*(text/html|application/xhtml\+xml)
RewriteCond %{HTTP_ACCEPT} text/html [OR]
RewriteCond %{HTTP_ACCEPT} application/xhtml\+xml [OR]
RewriteCond %{HTTP_USER_AGENT} ^Mozilla/.*
RewriteRule ^(.+)$ https://dgarijo.github.io/example/release/$1/index-en.html [R=303,L]

# Rewrite rule to serve Turtle serialization of a particular version
RewriteCond %{HTTP_ACCEPT} text/turtle [OR]
RewriteCond %{HTTP_ACCEPT} text/\* [OR]
RewriteCond %{HTTP_ACCEPT} \*/turtle 
RewriteRule ^(.+)$ https://dgarijo.github.io/example/release/$1/ontology.ttl [R=303,L]

# Rewrite rule for other non accepted formats
RewriteCond %{HTTP_ACCEPT} .+
RewriteRule ^(.*)$ https://dgarijo.github.io/example/release/1.0.1/406.html [R=406,L]

# Rewrite rule to serve the Turtle content from the vocabulary URI by default
RewriteRule ^$https://dgarijo.github.io/example/release/1.0.1/ontology.ttl [R=303,L]
\end{Verbatim}

%Note the redirections when the owl is being requested. If you have a slash vocabulary, you will have to follow the aforementioned W3C document for further instructions.

In order to test the redirection, the easiest way is just to paste the URI of the ontology in Prot\'eg\'e or in your browser and check that both load the right serialization. Another possibility is to use a \emph{curl} command,\footnote{\url{https://curl.haxx.se/}} e.g., to retrieve the Turtle serialization: 
\begin{verbatim}
curl -sH "Accept: text/turtle" -L https://w3id.org/example# 
\end{verbatim}
%(for requesting a Turtle serialization) or 
%\begin{verbatim}
%curl -sH "Accept: text/html" -L https://w3id.org/example#
%\end{verbatim}
%for HTML. 
If interested in a particular version, we can use its version IRI. For example, for the documentation of version 1.0.0:
\begin{verbatim}
curl -sH "Accept: text/html" -L https://w3id.org/example/1.0.0 
\end{verbatim}

%\textcolor{red}{Validating: Show CURL commands, it's the easiest way! Careful with https}

\subsection{Making an Ontology Findable on the Web}\label{sec:findable}
Once an ontology is published in the Web, the next step is to ensure it can be easily found by others. There are three main activities that can help the visibility of an ontology:

\begin{enumerate}
\item \textbf{Register the namespace prefix}: prefix.cc,\footnote{\url{http://prefix.cc}} is a crowdsourced registry where users can vote the most popular URL for a given prefix.

\item \textbf{Register the ontology}: There are a number of existing metadata registries that can be used for browsing existing ontologies \cite{vandenbussche_linked_2016}\cite{whetzel_bioportal:_2011}. Our recommendation is to look first for domain-specific registries (e.g., Bioportal \cite{whetzel_bioportal:_2011} in the biomedical domain, Agroportal \cite{JONQUET2018126} in Agronomy, etc.) commonly used by the target community of interest. When domain-specific registries do not exist, we suggest registering the ontology in a domain-generic metadata registry, such as Linked Open Vocabularies \cite{vandenbussche_linked_2016} (which has a manual curation process to ensure that minimum metadata is provided) or FAIRsharing.\footnote{\url{https://fairsharing.org/standards/}}

\item \textbf{In-document annotations} to help crawlers understand the metadata of the ontology when publishing it on the Web. For instance, annotations can be added in your HTML documentation through JSON-LD snippets:\footnote{\url{https://www.w3.org/TR/json-ld11/}}
\end{enumerate}

\begin{Verbatim}[frame=single, fontsize=\scriptsize]
<!-- Annotations for the example ontology -->
<script type="application/ld+json">{
   "@context":"http://schema.org",
   "@type":"WebPage",
   "url":"https://w3id.org/example",
   "name":"The example ontology", 
   "datePublished":"4-2-2020", 
   "version":"1.0.1", 
   "license":"http://creativecommons.org/licenses/by/2.0/", 
   "author":[{"@type":"Person","name":"Daniel Garijo"},
             {"@type":"Person","name":"Maria Poveda"}],
}</script>
\end{Verbatim}

\section{Ontology Documentation and Publication Frameworks}\label{sec:frameworks}
%\textcolor{red}{[Dani: @Maria, draft si quieres con lo basico]}  \maria{Done :-)}

%\textcolor{red}{[Dani: I think it would be nice to end up with frameworks and tools to generate some of these things. For example, OnToology, VoCol, others? }

%\textcolor{red}{[Dani: WebVowl, WIDOCO, LODE, Parrot... are tools for sections and we should mentione them wherever necessary.}

%During last decades, 
Semantic Web researchers and practitioners have developed methods and tools for easing  ontology engineering, development, publication and exploitation. A number these tools have already been mentioned in the corresponding sections of this chapter, however, they are not always integrated as part of an end-to-end framework, and researchers have to use them separately. 
%Some well-known ontology editors are Prot\'eg\'e\ \cite{musen_protege_2015}\url and TopBraid Composer.\footnote{\url{https://www.topquadrant.com/products/topbraid-composer/}} These editors are complemented by a number of plug-ins, open-source for the former and in some cases commercial for the later, to support numerous ontology development activities as testing, query support, data transformation, reasoners, etc. 

More recently, frameworks inspired in the software continuous integration practices have arisen to support ontology engineering activities in collaborative environments. These frameworks offer end-to-end solutions that support ontology engineers documenting, visualizing, testing and publishing their ontologies; and we recommend them as an entry point to adopt some of the practices described in this chapter.  One example is OnToology \cite{alobaid2018automating}, a web application\footnote{\url{http://ontoology.linkeddata.es/}} that orchestrates ontology documentation, %(including documentation for individual ontologies, diagrams generation and portal generation for set of ontologies)
evaluation and publication on the Web with permanent URLs. %This application does not need installation process nor dedicated servers and support ontologies stored in GitHub repositories. 
Another similar approach is VoCol \cite{DBLP:conf/ekaw/HalilajPG0ACL16}, which provides ontology developers with feedback on syntax and other errors and gives access to a human-readable presentation of a target ontology. %VoCol deployment should be done in developers servers and is also based on git environments. 
Finally, PoolParty \cite{schandl2010poolparty} is a commercial solution that also includes publication of thesauri, taxonomies and ontology management among other features.

\section{Conclusions}\label{sec:conclusions}
In this chapter we have described implementation guidelines and recommendations for making ontologies findable (through metadata registries and annotations); accessible (through good practices in URI design and content negotiation), interoperable (showing how to serve ontologies in different standard serializations) and reusable (by describing the metadata and diagram guidelines needed for proper understanding) on the Web while following the Linked Data principles. A distinct feature of our guidelines is that we have illustrated how to carry out our recommendations with an example ontology and pointers to usable tools developed by the Semantic Web community in the last decade. Our recommendations reflect years of experience on ontology engineering and also summarize community discussions for ontology design and publication. Hence, we believe these guidelines are a comprehensive starting point for ontology developers who aim to make their ontologies FAIR and available on the Web.

\bibliographystyle{splncs04}

\bibliography{bibliography}

\begin{thebibliography}{10}
\providecommand{\url}[1]{\texttt{#1}}
\providecommand{\urlprefix}{URL }
\providecommand{\doi}[1]{https://doi.org/#1}

\bibitem{alobaid2018automating}
Alobaid, A., Garijo, D., Poveda-Villal{\'o}n, M., Santana-Perez, I.,
  Fern{\'a}ndez-Izquierdo, A., Corcho, O.: {Automating ontology engineering
  support activities with OnToology}. Journal of Web Semantics  (2018)

\bibitem{alonso_current_2012}
Alonso, C.T., Berrueta, D., Polo, L., Fernández, S.: Current practices and
  perspectives for metadata on web ontologies and rules. International Journal
  of Metadata, Semantics and Ontologies  \textbf{7}(2), ~93 (2012)

\bibitem{damato_widoco:_2017}
Garijo, D.: {WIDOCO}: {A} {Wizard} for {Documenting} {Ontologies}. In: The
  {Semantic} {Web} – {ISWC} 2017, vol. 10588, pp. 94--102. Cham (2017)

\bibitem{haase2009d}
Haase, P., Brockmans, S., Palma, R., Euzenat, J., d'Aquin, M.: D1.1.2 updated
  version of the networked ontology model. Tech. rep., Universit{\"{a}}t
  Karlsruhe (2009), {NeOn Project}. http://www. neon-project. org

\bibitem{DBLP:conf/ekaw/HalilajPG0ACL16}
Halilaj, L., Petersen, N., Grangel{-}Gonz{\'{a}}lez, I., Lange, C., Auer, S.,
  Coskun, G., Lohmann, S.: Vocol: An integrated environment to support
  version-controlled vocabulary development. In: 20th International Conference
  on Knowledge Engineering and Knowledge Management - {EKAW} 2016, Bologna,
  Italy. Lecture Notes in Computer Science, vol. 10024, pp. 303--319 (2016)

\bibitem{janowicz_five_2014}
Janowicz, K., Hitzler, P., Adams, B., Kolas, D., Vardeman~II, C.: Five stars of
  {Linked} {Data} vocabulary use. Semantic Web  \textbf{5}(3),  173--176 (2014)

\bibitem{JONQUET2018126}
Jonquet, C., Toulet, A., Arnaud, E., Aubin, S., Yeumo, E.D., Emonet, V.,
  Graybeal, J., Laporte, M.A., Musen, M.A., Pesce, V., Larmande, P.:
  Agroportal: A vocabulary and ontology repository for agronomy. Computers and
  Electronics in Agriculture  \textbf{144},  126 -- 143 (2018)

\bibitem{le_franc_yann_2020_3707985}
Le~Franc, Y., Parland-von Essen, J., Bonino, L., Lehväslaiho, H., Coen, G.,
  Staiger, C.: D2.2 fair semantics: First recommendations (Mar 2020),
  \url{https://doi.org/10.5281/zenodo.3707985}

\bibitem{WebVOWL}
Lohmann, S., Link, V., Marbach, E., Negru, S.: {WebVOWL}: Web-based
  visualization of ontologies. In: Proceedings of EKAW 2014 Satellite Events.
  LNAI, vol.~8982, pp. 154--158. Springer (2015)

\bibitem{VOWL2}
Lohmann, S., Negru, S., Haag, F., Ertl, T.: {VOWL 2}: User-oriented
  visualization of ontologies. In: Proceedings of the 19th International
  Conference on Knowledge Engineering and Knowledge Management (EKAW~'14).
  LNAI, vol.~8876, pp. 266--281. Springer (2014)

\bibitem{musen_protege_2015}
Musen, M.A.: The {Prot\'eg\'e} project: a look back and a look forward. AI
  Matters  \textbf{1}(4),  4--12 (Jun 2015)

\bibitem{lode}
Peroni, S., Shotton, D., Vitali, F.: The {Live} {OWL} {Documentation}
  {Environment}: {A} {Tool} for the {Automatic} {Generation} of {Ontology}
  {Documentation}. In: Knowledge {Engineering} and {Knowledge} {Management},
  vol.~7603, pp. 398--412. Springer Berlin Heidelberg, Berlin, Heidelberg
  (2012)

\bibitem{schandl2010poolparty}
Schandl, T., Blumauer, A.: Poolparty: Skos thesaurus management utilizing
  linked data. In: Extended Semantic Web Conference. pp. 421--425. Springer
  (2010)

\bibitem{suarez-figueroa_neon_2010}
Suárez-Figueroa, M.C.: {NeOn} {Methodology} for building ontology networks:
  specification, scheduling and reuse. Ph.D. thesis, Facultad de Informatica,
  Universidad Politécnica de Madrid (2010)

\bibitem{vandenbussche_linked_2016}
Vandenbussche, P.Y., Atemezing, G.A., Poveda-Villalón, M., Vatant, B.: Linked
  {Open} {Vocabularies} ({LOV}): {A} gateway to reusable semantic vocabularies
  on the {Web}. Semantic Web  \textbf{8}(3),  437--452 (Jan 2017)

\bibitem{vrandecic_wikidata_2014}
Vrandečić, D., Krötzsch, M.: Wikidata: a free collaborative knowledgebase.
  Communications of the ACM  \textbf{57}(10),  78--85 (Sep 2014)

\bibitem{whetzel_bioportal:_2011}
Whetzel, P.L., Noy, N.F., Shah, N.H., Alexander, P.R., Nyulas, C., Tudorache,
  T., Musen, M.A.: {BioPortal}: enhanced functionality via new {Web} services
  from the {National} {Center} for {Biomedical} {Ontology} to access and use
  ontologies in software applications. Nucleic Acids Research
  \textbf{39}(suppl),  W541--W545 (Jul 2011)

\bibitem{wilkinson_fair_2016}
Wilkinson, M.D., Dumontier, M., Aalbersberg, I.J., Appleton, G., Axton, M.,
  Baak, A., Blomberg, N., et.al.: The {FAIR} {Guiding} {Principles} for
  scientific data management and stewardship. Scientific Data  \textbf{3},
  160018 (Mar 2016)

\end{thebibliography}

\end{document}